\documentclass[runningheads]{llncs}

\usepackage[T1]{fontenc}
\usepackage{graphicx}
\usepackage{subcaption}
\usepackage{color}
\usepackage{textcomp}
\usepackage{csquotes}
\usepackage{hyperref}
\usepackage[most]{tcolorbox}
\tcbuselibrary{listingsutf8}

\newtcolorbox{justifiedlisting}[1][]{%
  colback=gray!10,
  colframe=black!50,
  listing only,
  listing options={
    basicstyle=\small\ttfamily,
    breaklines=true,
    breakatwhitespace=true,
    columns=fullflexible,
    postbreak=\mbox{\textcolor{red}{$\hookrightarrow$}\space},
    #1
  },
  boxrule=0.5pt,
  arc=4pt,
  outer arc=4pt,
  enhanced,
  sharp corners,
}

\usepackage{caption} 
\usepackage{tabularx}
\usepackage{float}
\usepackage{array}  
\usepackage{booktabs} 
\usepackage{multirow}
\usepackage{makecell}
\usepackage{arydshln} 
\usepackage{todonotes}
\usepackage{eurosym}

\usepackage{siunitx}
\begin{document}
\title{Exploring LLM-generated Culture-specific Affective Human-Robot Tactile Interaction}

\titlerunning{DExploring LLM-generated Culture-specific Affective Human-Robot Tactile Interaction}

\author{Qiaoqiao Ren \and
Tony Belpaeme}
\authorrunning{Q. Ren et al.}
%
%
%
\institute{Faculty of Engineering and Architecture, IDLab-AIRO, Ghent University -- imec, Technologiepark 126, 9052 Gent, Belgium\\
\email{{Qiaoqiao.Ren}@ugent.be}}
\maketitle              
\begin{abstract}

As large language models (LLMs) become increasingly integrated into robotic systems, their potential to generate socially and culturally appropriate affective touch remains largely unexplored. This study investigates whether LLMs—specifically GPT-3.5, GPT-4, and GPT-4o—can generate culturally adaptive tactile behaviours to convey emotions in human–robot interaction. We produced text-based touch descriptions for 12 distinct emotions across three cultural contexts (Chinese, Belgian, and unspecified), and examined their interpretability in both robot-to-human and human-to-robot scenarios. A total of 90 participants (36 Chinese, 36 Belgian, and 18 culturally unspecified) evaluated these LLM-generated tactile behaviours for emotional decoding and perceived appropriateness. Results reveal that: (1) under matched cultural conditions, participants successfully decoded six out of twelve emotions—mainly socially oriented emotions such as love and Ekman emotions such as anger, however, self-focused emotions like pride and embarrassment were more difficult to interpret; (2) tactile behaviours were perceived as more appropriate when directed from human to robot than from robot to human, revealing an asymmetry in social expectations based on interaction roles; (3) behaviours interpreted as aggressive (e.g., anger), overly intimate (e.g., love), or emotionally ambiguous (i.e., not clearly decodable) were significantly more likely to be rated as inappropriate; and (4) cultural mismatches reduced decoding accuracy and increased the likelihood of behaviours being judged as inappropriate.

\keywords{Tactile interaction  \and affective computing \and emotion decoding
\and cultural difference}
\end{abstract}
\section{Introduction}

Tactile interaction plays a fundamental role in human social communication, providing an essential channel for conveying emotions, intentions, and nuanced social cues \cite{chen2024emotions}. As robotic systems become increasingly integrated into everyday human environments, the capacity for robots to understand and utilise tactile communication in culturally and socially appropriate ways is becoming crucial for enhancing the quality of human-robot interactions \cite{lim2021social}. However, generating emotionally expressive and contextually suitable tactile behaviours poses significant challenges, primarily due to the complexity of human touch, which varies greatly with cultural contexts and the relationships between interacting partners \cite{gallace2010science} \cite{sorokowska2021affective}. Current approaches in robotic tactile interaction often rely on predefined gestures or limited behavioural repertoires, lacking the flexibility required to adapt to diverse social contexts and cultural sensitivities.

Recent developments in artificial intelligence—particularly in large language models (LLMs) such as GPT-4, GPT-4o, and GPT-3.5—offer promising avenues for overcoming limitations in human-robot interaction \cite{zhang2025generative}. Although LLMs inherently lack physical embodiment and direct sensory experiences \cite{wang2024large}, their extensive training on vast and diverse textual corpora endows them with implicit knowledge of social and cultural nuances \cite{pawar2024survey}. Prior research has demonstrated the use of LLMs in generating the robot's expressive behaviours. Furthermore, recent studies have explored how LLMs can guide robots in generating social behaviours \cite{mahadevan2024generative}, including affective tactile interactions via devices such as vibration sleeves \cite{ren2025touched}. 


While LLMs show great potential in driving robotic applications, human-robot tactile interactions are inherently contextual and culturally situated \cite{cekaite2020towards}. In addition, one key factor shaping the interpretation of tactile interaction is the role of the robot—whether it engages in active or passive touch. For example, studies on robot-initiated (active) touch have shown that the robot’s perceived intent significantly influences human responses. Participants tend to prefer instrumental touch (e.g., touch used to assist or guide) over affective touch, which may feel socially ambiguous or intrusive \cite{willemse2019social}. On the other hand, passive touch, where the robot is designed to be touched, has also been explored in systems such as the Haptic Creature \cite{yohanan2012role}, which was reported to induce calmness and happiness in users. Similarly, recent work has demonstrated that robot-mediated passive touch can help reduce stress levels during risk-taking tasks \cite{ren2024tactile}.

However, most of these studies have focused on positive social touch. In contrast, affective tactile interaction is more complex, as it also involves negative or ambiguous emotions, which can be harder for robots to interpret or express appropriately. Therefore, it is essential for robots to develop an awareness of what types of touch are appropriate, and when, to avoid crossing social or cultural boundaries. As LLMs become more integrated into embodied systems, their ability to interpret context, understand interaction roles, and generate culturally appropriate tactile behaviours will be critical for safe and meaningful human–robot touch interaction.


To address this gap, this study investigates the ability of large language models (LLMs) to generate culturally sensitive and socially appropriate tactile behaviour descriptions intended to convey distinct emotions. Specifically, we explore the following research questions: 1) How accurately can LLM-generated tactile descriptions be decoded across different cultural contexts, and which types of emotions are more or less easily recognised? 2) How does the direction of interaction (human-to-robot vs. robot-to-human) influence the perception of social roles and appropriateness in tactile communication? 3) Which decoded emotions are most frequently perceived as inappropriate, and do they pose potential risks for social misunderstanding? 4) Can LLMs adapt to different cultural contexts when generating text-based affective tactile behaviours?

\section{Materials \& Methods}
\label{sec2:Materials}

In this study, participants were asked to interpret descriptions of affective tactile behaviours. Participants either had a Chinese background, a Belgian background, or an unspecified cultural background. The descriptions were generated by large language models (LLMs), using culturally tailored prompts (see Section~\ref{lst:prompt}). The following sections describe the equipment, setup, and data acquisition procedures.


\vspace{-1em}

\subsection{Emotions}

Inspired by \cite{hertenstein2006touch}, we selected a diverse set of emotions to examine tactile emotional communication. These included: (a) six Ekman emotions—anger, fear, happiness, sadness, disgust, and surprise; (b) three prosocial emotions associated with cooperation and altruism—love, gratitude, and sympathy; and (c) three self-focused emotions—embarrassment, pride, and envy.

\vspace{-1em}

\subsection{Prompt}
\label{lst:prompt}



Participants were shown textual descriptions of affective tactile behaviours, which were generated by an LLM using the following prompt: ``You are an encoder (human or humanoid robot) tasked with expressing one of 12 emotions solely through touch to a decoder/receiver (humanoid robot or human) within a specific cultural context (Chinese, Belgian, or None). The touch should be applied to the bare arm, from the elbow to the hand. Provide 10 distinct tactile behaviours to express the emotion. For each, describe the touch behaviour, touch intensity, its rhythm or timing, and how cultural context is considered. Focus solely on tactile behaviours, ensuring the behaviours are contextually and culturally appropriate.'' The detailed prompt is available on GitHub\footnote{\url{https://github.com/qiaoqiao2323/LLM_touch_culture/tree/main}}. Here is an example of generated tactile behaviour for anger; the touch expression generated for the Belgian culture context is \textit{a swift slap to the back of the hand}. In addition, the Chinese culture context suggests \textit{a quick drag down the arm, firm and assertive.}


\vspace{-1em}

\subsection{Experimental design}


We recruited 90 participants (57 identifying as male, 32 as female, and one preferring not to report gender), ages 19 to 51. The sample included 36 Chinese participants, 36 Belgian participants, and 18 participants without a specified cultural context. The study adhered to the ethical procedures of \emph{Ghent University}, and all participants provided informed consent. Participants were asked to decode emotions from LLM-generated tactile behaviours. For stimulus creation, we generated 18 unique series of tactile behaviours, each series containing 3 tactile behaviours for each of 12 target emotions. The tactile behaviours were derived from the following combination of experimental conditions:

\begin{equation}
\text{Interaction Direction (2)} \times \text{Culture (3)} \times \text{LLM Type (3)} =  \text{18 series}
\end{equation}

\noindent where:
\begin{itemize}
    \item \textbf{Interaction Direction}: \{Robot-to-Human, Human-to-Robot\}
    \item \textbf{Cultural Context}: \{Chinese, Belgian, No specified context\}
    \item \textbf{LLM Type}: \{GPT-4, GPT-4o, GPT-3.5-turbo\}
\end{itemize}

Following the evaluation process proposed by \cite{janssens2024integrating}, each series included a total of 36 tactile behaviours (3 behaviours for each of the 12 emotions) and each series was rated by three participants, with the constraint that no participant evaluated more than one series and no tactile behaviour appeared more than once within a series.

\subsection{Procedure}

Each participant completed a questionnaire for each of the 36 tactile behaviours. Specifically, for each of the 12 target emotions, we generated three different tactile behaviours. For each tactile behaviour, participants were provided with contextual information describing the direction of emotional expression: either the robot was conveying an emotion to a human (\textit{robot-to-human}) or a human was conveying an emotion to a robot (\textit{human-to-robot}). This directionality was made explicit at the beginning of each block or stimulus description.

To explore cultural adaptation in tactile behaviours, these behaviours were generated by LLMs under three cultural prompt conditions: Chinese, Belgian, and culturally unspecified (none). Participants were either matched or mismatched with these cultural conditions. Specifically, 18 Chinese participants received tactile behaviours generated for the Chinese cultural context (Chi-Chi), while another 18 Chinese participants received behaviours generated for the Belgian context (Bel-Chi). Similarly, 18 Belgian participants experienced tactile behaviours tailored to the Belgian context (Bel-Bel), and another 18 Belgian participants received those generated for the Chinese context (Chi-Bel). Additionally, 18 participants without a clearly defined cultural background received tactile behaviours generated under the culturally unspecified condition. The presentation of the descriptions was balanced across participants. Participants were instructed to imagine the described interaction and then evaluate each tactile behaviour accordingly. After each behaviour, they responded to two questions:

\begin{enumerate}
    \item \textbf{Emotion Decoding:} ``Please choose the term that best describes what this [robot or human] is communicating.'' Followed by the forced-choice methodology \cite{hertenstein2006touch}, thirteen forced-choice options were provided: \textit{anger, disgust, fear, happiness, sadness, surprise, sympathy, embarrassment, love, envy, pride, gratitude}, and \textit{none of these terms is correct}. 
    
    \item \textbf{Appropriateness Judgment:} “Do you think this tactile behaviour is appropriate?” Participants were instructed to select “Appropriate” if the behaviour could be considered appropriate for one of any scenarios in the described interaction context, and “Inappropriate” otherwise. If participants were uncertain, they could select the “Maybe” option.
\end{enumerate}


\section{Results and analysis}
\label{sec::data_analysis}

\subsection{Culture decoding results analysis}

We evaluated the decoding accuracy of participants across three cultural groups - Chinese, Belgian, and culturally unspecified (None)—for 12 emotions expressed through LLM-generated affective tactile behaviours. Each participant rated 54 stimuli, we got 3240 measurements in total. We performed one-sided binomial tests against the chance level ($1/13 \approx 7.7\%$) to assess whether emotion decoding results are significantly above chance. Bonferroni correction was applied to account for multiple comparisons.

\vspace{-1em}

\subsubsection{Chi-Chi Participants decoding results}
\label{chi-chi}

As shown in Table~\ref{tab:emotion_tactile}, 18 Chinese participants achieved significantly above-chance decoding for \textbf{anger} (50\%, $p_\text{adj} < .001$), \textbf{fear} (37.0\%, $p_\text{adj} < .001$), \textbf{gratitude} (27.8\%, $p_\text{adj} < .001$), \textbf{love} (50\%, $p_\text{adj} < .001$), \textbf{surprise} (24.1\%, $p_\text{adj} < .01$), and \textbf{sympathy} (40.7\%, $p_\text{adj} < .001$). Emotions such as \textbf{embarrassment}, \textbf{envy}, \textbf{pride}, and \textbf{happiness} did not exceed the chance threshold. The most frequently misclassified emotion was \textbf{love}, which was commonly mistaken for \textbf{gratitude} and \textbf{sympathy}, especially for emotions like \textbf{happiness}, \textbf{sadness}, and \textbf{disgust}. The results further demonstrate that although participants struggled to interpret all the LLM-generated affective tactile behaviours, they were still able to decode 6 out of 12 emotions at levels significantly above chance. Interestingly, previous studies on human-to-human tactile emotion decoding also found that individuals could reliably recognise six emotions—primarily prosocial emotions and three of Ekman's basic emotions: anger, fear, and disgust—while self-focused emotions remained difficult to identify. This suggests that LLMs are capable of generating meaningful affective tactile behaviours from text, with potential applications in guiding robots to perform culturally sensitive affective touch, which is comparable to the human encoder. Prior work has shown that robots can already be driven by LLMs to execute behaviours and facilitate mediated tactile interactions. Given that tactile communication is highly dependent on context and cultural norms, this study represents an important first step toward understanding how LLMs can support culturally adaptive affective touch, which could be a guide for future LLM-driven tactile interaction design.

\vspace{-1em}

\subsubsection{Chi-Bel Participants decoding results}

In the Chi-Bel condition, 18 Belgian participants were asked to decode the same set of tactile behaviours generated by the LLM for the Chinese cultural context (see Subsection.~\ref{chi-chi}). The results showed significantly poorer performance compared to the Chi-Chi group. Among the 12 target emotions, only Love was successfully decoded above chance level ($p_\text{adj} < .01$). Specifically, participants in the Chi-Bel group correctly identified 1 out of 12 emotions, while participants in the Chi-Chi group decoded 6 out of 12 emotions successfully based on tactile behaviour alone. This discrepancy suggests that the LLM-generated tactile expressions may be more aligned with Chinese cultural norms, making them less interpretable for individuals from a Belgian cultural background.

\subsubsection{Bel-Bel Participants decoding results}

\label{bel-bel}

Belgian participants also demonstrated above-chance accuracy for several emotions: \textbf{anger} (59.3\%, $p_\text{adj} < .001$), \textbf{fear} (31.5\%, $p_\text{adj} < .001$), \textbf{gratitude} (20.4\%, $p_\text{adj} < .05$), \textbf{love} (35.2\%, $p_\text{adj} < .001$), \textbf{surprise} (22.2\%, $p_\text{adj} < .01$), and \textbf{sympathy} (37.0\%, $p_\text{adj} < .001$). However, recognition of self-focused emotions (\textbf{pride}, \textbf{embarrassment}, \textbf{envy}) and basic emotions like \textbf{happiness} and \textbf{disgust} remained at or near chance level. Misclassification patterns showed \textbf{happiness} was often confused with \textbf{love}, and \textbf{fear} was frequently confused with \textbf{surprise}.

\vspace{-1em}

\subsubsection{Bel-Chi Participants decoding results}

In the Bel-Chi group, 18 Chinese participants were asked to decode the same set of tactile behaviours generated by the LLM for the Belgian cultural context, identical to those used in the Bel-Bel group. The results indicate that Chinese participants were able to successfully decode anger and love at rates significantly above the chance level. This finding suggests that tactile behaviours generated for the Belgian cultural context may be aligned with Belgian norms and less suitable for Chinese cultural interpretations.



\begin{figure}[t]
    \centering
        \includegraphics[height=5cm]{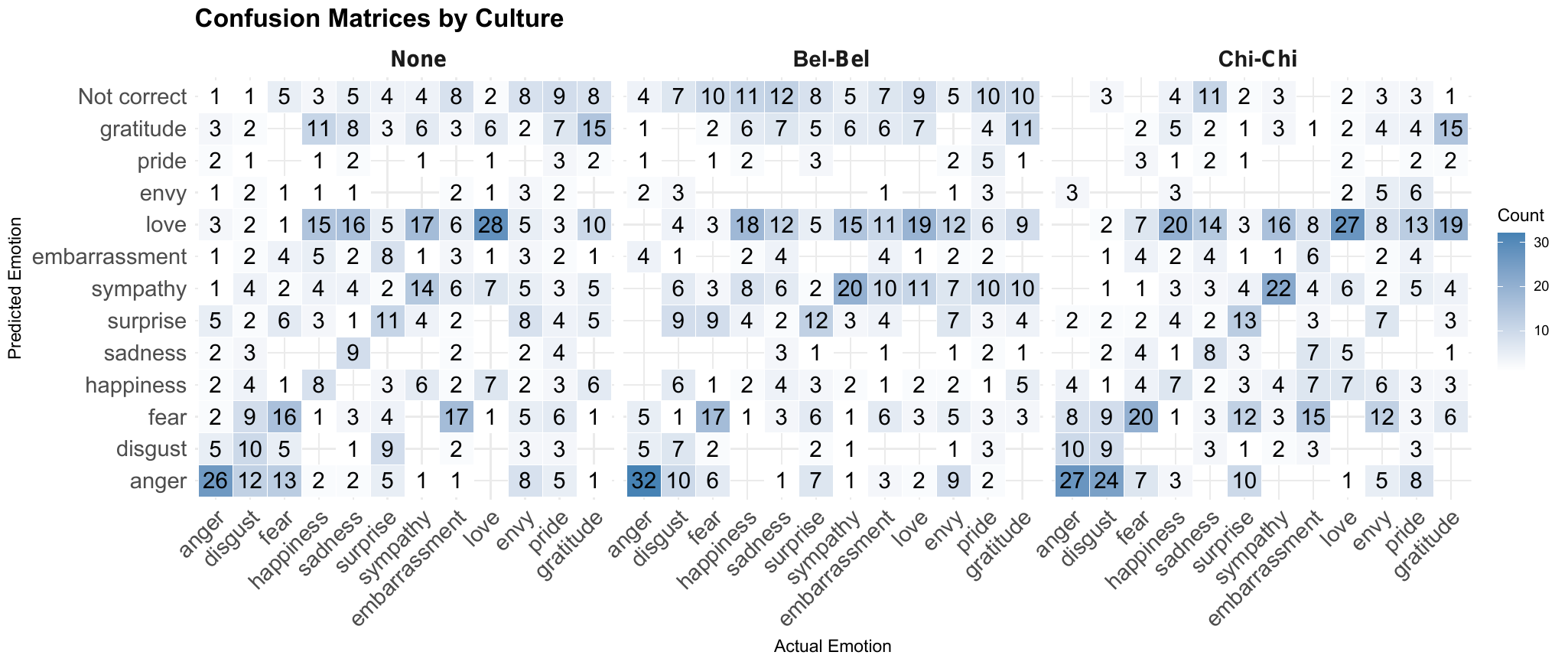}
        \vspace{-1.2em} 
        \includegraphics[height=5.cm]{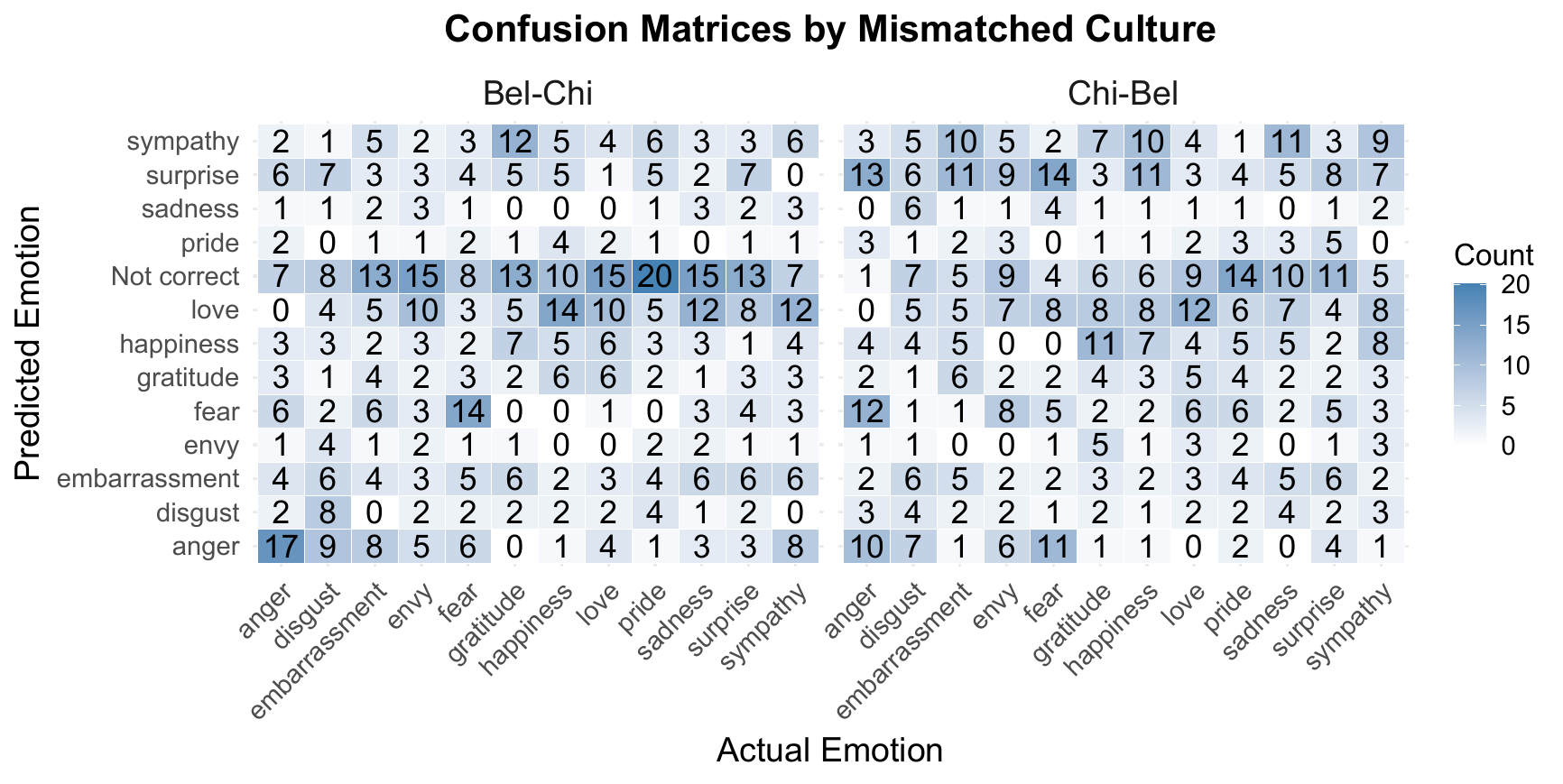}
    \caption{Emotions across cultural groups (top) and for misinterpreted or inappropriate tactile responses (bottom).}
    \label{fig::confusion_matrix}
\end{figure}

\vspace{-1em}

\subsubsection{Participants from non-specified culture}

Across none cultural contexts, decoding performance remained significantly above chance for \textbf{anger} (48.1\%, $p_\text{adj} < .001$), \textbf{fear} (29.6\%, $p_\text{adj} < .001$), \textbf{gratitude} (27.8\%, $p_\text{adj} < .001$), \textbf{love} (51.9\%, $p_\text{adj} < .001$), \textbf{surprise} (20.4\%, $p_\text{adj} < .05$), and \textbf{sympathy} (25.9\%, $p_\text{adj} < .001$). This confirms the overall reliability of LLM-generated tactile behaviours in conveying these emotions. In contrast, emotions like \textbf{envy}, \textbf{pride}, and \textbf{embarrassment} were consistently difficult to recognise across all groups. Misclassification patterns revealed recurring confusion among closely related emotions, particularly between prosocial emotions such as \textbf{love}, \textbf{sympathy}, and \textbf{gratitude}.



\begin{table*}\footnotesize 
\setlength{\abovecaptionskip}{0.0cm}   
	\setlength{\belowcaptionskip}{-0cm}  
	\renewcommand\tabcolsep{2.0pt} 
	\centering
	\caption{MDA refers to Matched-culture Decoding Accuracy, while MIDA refers to Mismatched-culture Decoding Accuracy (** at p < 0.01). FCE represent the most Frequently Chosen Emotions in the decoding of each target emotion and its corresponding frequency ($f$).}
	\begin{tabular}
	{
	p{2cm}<{\centering} 
 p{1.5cm}<{\centering} 
	 p{1.9cm}<{\centering} 
  p{1.5cm}<{\centering} |
	p{1.8cm}<{\centering}
 	p{2.5cm}<{\centering}
     	p{1.5cm}<{\centering}
	} 
        \hline
    
        \textbf{Emotion} & \textbf{Culture} & \textbf{FCE/$f$} &  \textbf{MDA(\%)} &  \textbf{Culture} &  \textbf{FCE/$f$} &  \textbf{MIDA(\%)} \\
        \multicolumn{2}{l}{\textbf{Ekman’s emotions}} & & &\\
        \hline

        \multirow{3}{*}{Anger} & Chi-Chi & anger/27 & {50**} & Chi-Bel & anger/17 & {18.5} \\
          & Bel-Bel & anger/32 & {59.3**}  & Bel-Chi & surprise/13 & {31.5**}\\
          & None & anger/26 &  {48.1**} & / & / & / \\
        \hline
        \multirow{3}{*}{Fear} & Chi-Chi & fear/20 & {37.0**}& Chi-Bel & fear/14 & {9.3} \\
          & Bel-Bel & fear/17 & {31.5**}  &  Bel-Chi & surprise/14 & {25.9**} \\
          & None & fear/16 & {29.6**} & / & / & / \\
        \hline
        \multirow{3}{*}{Happiness} & Chi-Chi & love/20 & {13.0}& Chi-Bel &  love/14 & {13.0}\\
          & Bel-Bel & love/18 & {3.7} & Bel-Chi & surprise/11 & {9.3} \\
          & None & love/15 & {14.8}  & / & / & / \\
        \hline
        \multirow{3}{*}{Sadness}& Chi-Chi & love/14 & {14.8} & Chi-Bel & None correct/15 & {0} \\
          & Bel-Bel & love/12 & {5.6} & Bel-Chi & sympathy/11 & {5.6} \\
          & None & love/16 & {16.7}  & / & / & / \\
        \hline
        \multirow{3}{*}{Disgust} & Chi-Chi & anger/24 & {16.7} & Chi-Bel & anger/9 & {7.4}\\
          & Bel-Bel & anger/10 & {13.0} & Bel-Chi & anger/7 & {14.8} \\
          & None & anger/12 & {18.5}  & / & / & / \\
        \hline
        \multirow{3}{*}{Surprise}& Chi-Chi & fear/12 & {24.1**} & Chi-Bel & None correct/13 & {14.8}\\
          & Bel-Bel & surprise/12 & {22.2**}& Bel-Chi & None correct/11 & {13.0}\\
          & None & surprise/11 & {20.4*} & / & / & / \\
        \hline
        \multicolumn{2}{l}{\textbf{Self-focused emotions}} & & &  \\
        \multirow{3}{*}{Embarrassment} & Chi-Chi & fear/15 & {11.1}  & Chi-Bel & None correct/13 & {9.3} \\
          & Bel-Bel & love/11 & {7.4}  & Bel-Chi & love/12 & {7.4} \\
          & None & fear/17 & {5.6}  & / & / & / \\
        \hline
        \multirow{3}{*}{Envy} & Chi-Chi & fear/12m & {9.3}  & Chi-Bel & None correct/9 & {0} \\
          & Bel-Bel & love/12 & {1.9}  & Bel-Chi & disgust/5 & {3.7} \\
          & None & anger/8 & {5.6} & / & / & / \\
        \hline
        \multirow{3}{*}{Pride} & Chi-Chi & love/13 & {3.7}  & Chi-Bel & None correct/20 & {5.6} \\
          & Bel-Bel & None correct/10 & {9.3}  & Bel-Chi & None correct/14 & {1.9} \\
          & None & None correct/9 & {5.6}  & / & / & / \\
        \hline
        \multicolumn{2}{l}{\textbf{Prosocial emotions}} & & & \\
        \multirow{3}{*}{Love} & Chi-Chi & love/27 & {50**}  & Chi-Bel & None correct/15 & {22.2**}\\
          & Bel-Bel & love/19 & {35.2**}  & Bel-Chi & love/12 & {1.9} \\
          & None & love/28 & {51.9**} & / & / & / \\
        \hline
        \multirow{3}{*}{Gratitude} & Chi-Chi & love/19 & {27.8**}  & Chi-Bel & None correct/13 & {7.4} \\
          & Bel-Bel & gratitude/11 & {20.4*}  & Bel-Chi & happiness/11 & {3.7} \\
          & None & gratitude/15 & {27.8**}  & / & / & / \\
        \hline
        \multirow{3}{*}{Sympathy} & Chi-Chi & sympathy/22 & {40.7**}  & Chi-Bel & love/12 & {16.7} \\
          & Bel-Bel & sympathy/20 & {37.0**}  & Bel-Chi & sympathy/9 & {11.1} \\
          & None & love/17 & {25.9**}  & / & / & / \\
                  \hline

    \end{tabular}
    \caption{Decoding accuracy and confusion emotions. * p < .05, ** p < .01.}
    \label{tab:emotion_tactile}
\end{table*}

\subsection{Interaction direction}

Table~\ref{tab::direction} presents the number of responses categorised as “Appropriate,” “Inappropriate,” and “Maybe” for tactile behaviours in both the \textit{robot-to-human} and \textit{human-to-robot} conditions across different cultural groups.

To investigate cultural differences in the perception of tactile appropriateness, we conducted Chi-square goodness-of-fit tests within each cultural group (Chi-Chi, Chi-Bel, Bel-Bel, Bel-Chi, and All) to compare the frequencies of ‘appropriate’ and `inappropriate’ judgments. As shown in Table.~\ref{tab::direction}, the results revealed that all groups—except \textit{Bel-Chi}—judged significantly more tactile behaviours as appropriate than inappropriate, suggesting a general cultural acceptance of the stimuli. Specifically, significant effects were observed in the \textit{Chi-Chi} ($\chi^2 = 106.0$, $p < .001$), \textit{Bel-Bel} ($\chi^2 = 82.4$, $p < .01$), \textit{Chi-Bel} ($\chi^2 = 34.2$, $p < .01$), and \textit{None} ($\chi^2 = 124.0$, $p < .01$) groups. In contrast, the \textit{Bel-Chi} group did not show a significant preference for either response category ($\chi^2 = 3.0$, $p = .08$).

To examine between-group cultural mismatches, we conducted Chi-square tests of independence on appropriateness ratings between \textit{Chi-Chi} vs. \textit{Chi-Bel} and \textit{Bel-Bel} vs. \textit{Bel-Chi}. Significant group differences emerged in both comparisons. Notably, the \textit{Chi-Chi} group gave significantly more \textit{‘appropriate’} ratings ($\chi^2 = 5.20$, $p = .02$) and fewer \textit{‘inappropriate’} ratings ($\chi^2 = 12.0$, $p < .01$) than the \textit{Chi-Bel} group. Similarly, the \textit{Bel-Bel} group rated significantly more \textit{‘appropriate’} ($\chi^2 = 21.5$, $p < .01$) and fewer \textit{‘inappropriate’} behaviours ($\chi^2 = 21.6$, $p < .01$) than the \textit{Bel-Chi} group. These findings suggest that when LLMs generate tactile behaviours aligned with a specific cultural context, they may fail to adhere to the affective touch norms of other cultures. Consequently, participants from different cultural backgrounds are more likely to perceive such behaviours as socially inappropriate—an effect clearly reflected in the increased number of inappropriate ratings across mismatched groups.

To evaluate the influence of reciever identity (Human vs. Robot), we conducted Chi-square tests on appropriateness ratings. Results showed that human-to-robot interactions were significantly more associated with \textit{‘inappropriate’} judgments than robot-to-human interactions ($\chi^2 = 76.8$, $p < .001$), while robot-initiated expressions were more likely to be judged as \textit{‘appropriate’} ($\chi^2 = 18.8$, $p < .001$). Further analysis revealed that even when participants decoded the tactile behaviours as the same emotion, they showed greater caution in accepting robot-initiated touch toward humans than the reverse. Participants appeared more accepting of human-initiated emotional expressions toward robots.

Participants reported that many LLM-generated tactile behaviours were perceived as overly intimate or unnatural. While participants understood the intent to convey emotion, they noted that such gestures would rarely be used in real-life interactions. Discomfort was particularly high in the \textit{robot-to-human} condition, where participants reported greater unease imagining themselves as the receiver of such touch. In contrast, emotional expressions from human to robot—especially for certain emotions—were seen as more acceptable,.

These findings indicate a potential asymmetry in the LLMs’ capacity to model socially appropriate touch depending on the direction of interaction. LLMs appear to generate more plausible and acceptable tactile behaviours when simulating a human expressing emotion to a robot, possibly due to prompt design biases or a limited understanding of the social boundaries associated with human tactile reception.

\begin{table}\footnotesize 
\setlength{\abovecaptionskip}{0.0cm}   
	\setlength{\belowcaptionskip}{-0cm}  
	\renewcommand\tabcolsep{2.0pt} 
	\centering
	\caption{Appropriateness (number and its corresponding percentage) of the given tactile behaviour description.}
	\begin{tabular}
	{
	p{1.5cm}<{\centering} 
 p{1.9cm}<{\centering} 
	 p{1.9cm}<{\centering} 
  p{1.9cm}<{\centering}
	p{1.9cm}<{\centering}
	} 
\hline
    
     {Culture} & 
     {Reciever}  & 
     {Appropriate} & 
     {Inappropriate} & 
     {Maybe} \\

     \cline{1-5}

   \multirow{2}{*}{Chi-Chi} &
{Human} &
 {(170, 52.5\%)} &
 {(95, 29.3\%)} &
  {(59, 18.2\%)} \\

 & {Robot} & 
     {(193, 59.6\%}) & 
     {(39, 12.0\%)} & 
     {(92, 28.4\%)} \\

     \hline
   \multirow{2}{*}{Chi-Bel} &
  {Human} &
 {(144, 44.4\%)} &
 {(128, 39.5\%)} &
  {(52, 16.0\%)} \\

 & {Robot} & 
     {(177, 54.6\%)} & 
     {(61, 18.8\%)} & 
     {(86, 26.5\%)} \\

\hline
   \multirow{2}{*}{Bel-Bel} &
  {Human} &
 {(131, 40.4\%)} &
 {(95, 29.3\%)} &
  {(98, 30.2\%)} \\

 & {Robot} & 
     {(200, 61.7\%)} & 
     {(40, 12.3\%)} & 
     {(84, 25.9\%)} \\

    \hline
   \multirow{2}{*}{Bel-Chi} &
  {Human} &
 {(127, 39.2\%)} &
 {(112, 34.6\%)} &
  {(85, 26.2\%)} \\

 & {Robot} & 
     {(120, 37.0\%)} & 
     {(98, 30.2\%)} & 
     {(106, 32.7\%)} \\
     
     \hline
       \multirow{2}{*}{All} &
  {Human} &
 {(154, 47.5\%)} &
 {(87, 26.9\%)} &
  {(83, 25.6\%)} \\

 & {Robot} & 
     {(211, 65.1\%)} & 
     {(33, 10.2\%)} & 
     {(80, 24.7\%)} \\
  \hline

	\end{tabular}
	\label{tab::direction}
\end{table}

\vspace{-2em}

\subsection{Misclassification and Appropriateness Patterns}

\begin{figure}[t]
    \centering
        \includegraphics[height=5cm]{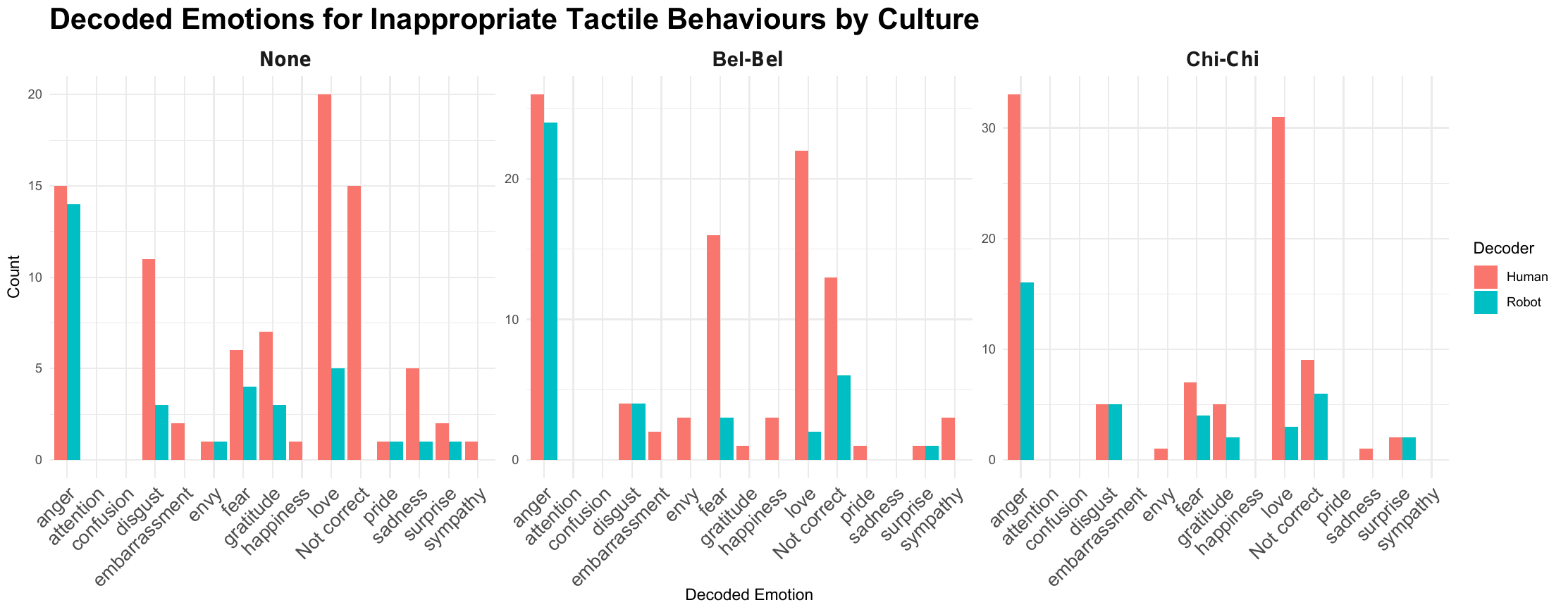}
        \vspace{-1.2em} 
        \includegraphics[height=5cm]{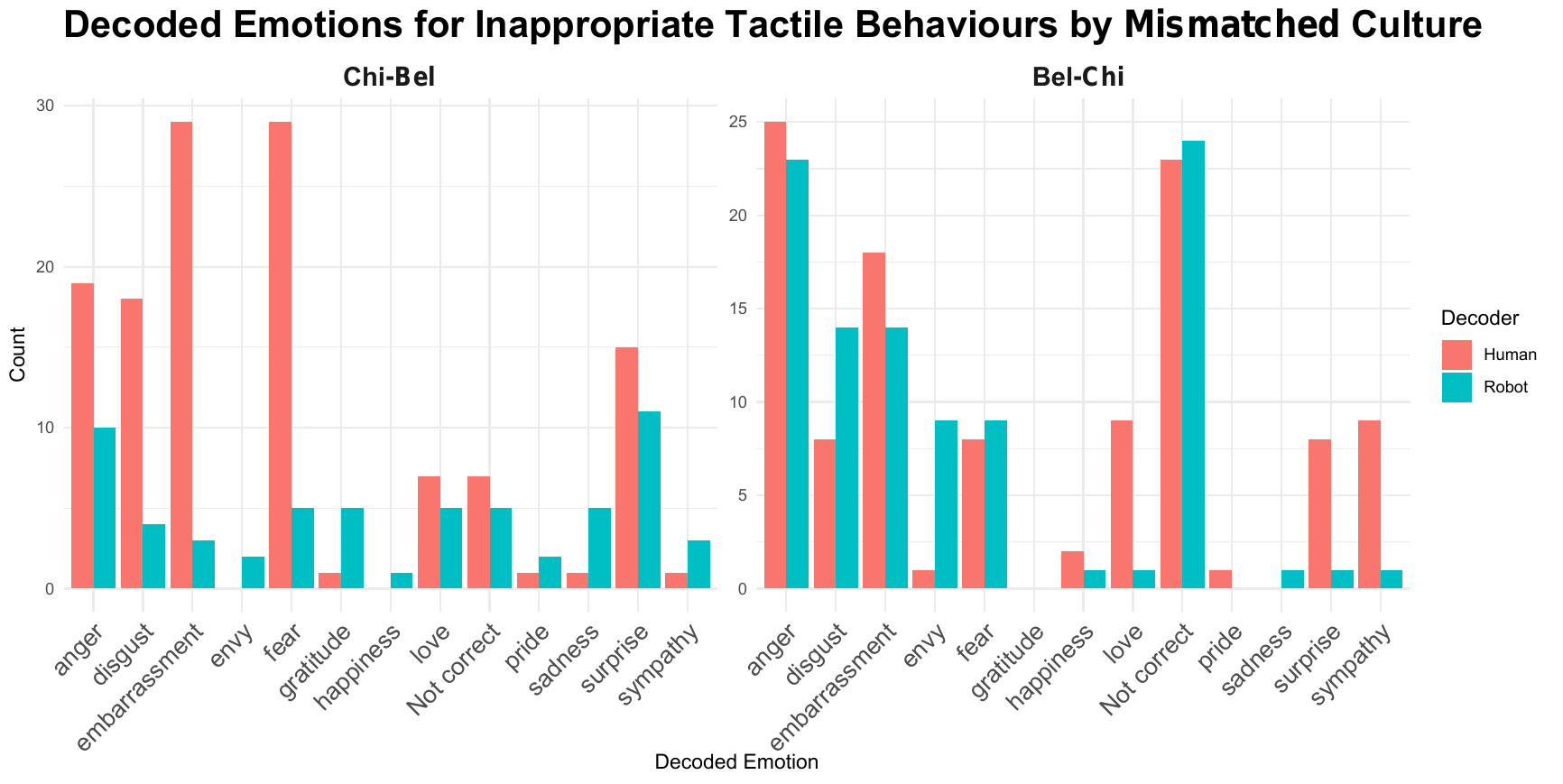}
    \caption{Emotions across cultural groups (top) and for misinterpreted or inappropriate tactile responses (bottom). Red bars indicate \textit{robot-to-human}, and blue bars indicate that \textit{human-to-robot}}.
    \label{fig:culture_combined}
\end{figure}

We further examined the most frequently chosen emotions for each target emotion, as shown in Table~\ref{tab:emotion_tactile} and Fig.\ref{fig::confusion_matrix}. Additionally, we investigated which tactile behaviours were perceived as inappropriate and analysed their corresponding decoded emotions. As illustrated in Fig.\ref{fig:culture_combined}, participants tended to associate certain negative or highly intimate emotions, such as anger and even love, with inappropriate tactile expressions. For example, one touch behaviour decoded as anger was described as "a quick, sharp pinch on the skin followed by an abrupt release", while a touch behaviour decoded as love is "lightly running fingertips along the human's arm, starting from the elbow and moving toward the hand in a soothing and gentle manner." In mismatched cultural conditions, a similar pattern emerged: participants were more likely to judge behaviours as inappropriate when they were associated with negative emotions or when the intended emotion could not be accurately decoded. Interestingly, some tactile behaviours interpreted as conveying love were also rated as inappropriate—likely due to their perceived level of physical intimacy. These findings highlight the social boundaries and cultural sensitivities surrounding affective touch, especially when generated by non-human agents.

For matched cultures, the misclassification data showed that participants tended to decode the intended emotions as socially oriented ones or the Ekman six basic emotions, such as love, anger and fear.  Notably, emotions like envy, pride, and embarrassment were frequently misinterpreted as prosocial emotions such as love, sympathy, or gratitude. This trend suggests that participants struggled to decode self-focused or socially complex emotions and instead defaulted to more familiar or socially desirable interpretations. Additionally, emotions decoded as negative, such as anger, were often rated as inappropriate. For mismatched cultures, they tend to decode tactile behaviours as none of the given emotions are correct. 


\subsection{Limitations}
This study has several limitations. First, participants decoded tactile behaviours based solely on textual descriptions, without any visual or physical representation. Providing visualisations or haptic feedback could improve ecological validity and decoding accuracy. Second, while we included both Chinese and Belgian participants, the findings may not generalise to other cultures with different touch norms. Third, to explore the expressive range of LLMs, we intentionally avoided placing strict moral or social constraints on the prompts. This design choice may have led to the generation of behaviours perceived as overly intimate or unrealistic. In addition, the results might be influenced by the education, the participants's attention and familiarity with the robot's tactile interaction. Moreover, all the questionnaires are in English, which could also limit the participants' interpretation as English is not the native language of Chinese and Belgian participants. Lastly, the interaction design was unidirectional, with LLMs serving only as encoders. Future work should explore bidirectional human-robot tactile interaction to better reflect real-world dynamics.

\section{Conclusion}
\label{sec::conclusion}

This study investigated the potential of large language models (LLMs) to generate culturally and socially appropriate tactile behaviours for conveying emotion in human-robot interaction. By examining participants from different cultural backgrounds, we identified key insights into how LLM-generated touch behaviours are perceived across emotion types and interaction directions. First, LLMs were generally successful at producing tactile behaviours that conveyed socially oriented emotions, such as anger, fear, love, and sympathy, within cultural contexts. However, self-focused emotions like embarrassment, pride, and envy were more difficult for participants to decode, highlighting inherent challenges in modelling these more introspective states. Second, interaction direction played a significant role in perceived appropriateness: participants rated behaviours as more acceptable when humans expressed emotion toward robots than when robots initiated touch. This asymmetry underscores the importance of contextual grounding and role expectations in human-robot touch dynamics. Third, our findings reveal that LLMs are capable of cultural adaptation when generating affective touch behaviours. When emotion descriptions were mismatched with a participant's cultural context, decoding performance declined, and behaviours were more frequently rated as inappropriate. This suggests that LLMs can incorporate cultural cues when guided properly, but also highlights the risks of cross-cultural misalignment.

Together, these results demonstrate the promise of LLMs in supporting affective tactile interaction, while also emphasising the critical need for cultural sensitivity and contextual awareness in the design of socially intelligent robots. This work represents an important step toward leveraging LLMs for culturally adaptive and emotionally expressive tactile interfaces in future human-robot interaction.

\bibliographystyle{unsrt}
\bibliography{references.bib}

\begin{thebibliography}{10}

\bibitem{chen2024emotions}
Xinyi Chen and Meng~Ting Zhang.
\newblock Emotions: Investigating the vital role of tactile interaction.
\newblock In {\em International Conference on Human-Computer Interaction}, pages 326--344. Springer, 2024.

\bibitem{lim2021social}
Velvetina Lim, Maki Rooksby, and Emily~S Cross.
\newblock Social robots on a global stage: establishing a role for culture during human--robot interaction.
\newblock {\em International Journal of Social Robotics}, 13(6):1307--1333, 2021.

\bibitem{gallace2010science}
Alberto Gallace and Charles Spence.
\newblock The science of interpersonal touch: an overview.
\newblock {\em Neuroscience \& Biobehavioral Reviews}, 34(2):246--259, 2010.

\bibitem{sorokowska2021affective}
Agnieszka Sorokowska, Supreet Saluja, Piotr Sorokowski, Tomasz Fr{\k{a}}ckowiak, Maciej Karwowski, Toivo Aavik, Grace Akello, Charlotte Alm, Naumana Amjad, Afifa Anjum, et~al.
\newblock Affective interpersonal touch in close relationships: A cross-cultural perspective.
\newblock {\em Personality and Social Psychology Bulletin}, 47(12):1705--1721, 2021.

\bibitem{zhang2025generative}
Kun Zhang, Peng Yun, Jun Cen, Junhao Cai, Didi Zhu, Hangjie Yuan, Chao Zhao, Tao Feng, Michael~Yu Wang, Qifeng Chen, et~al.
\newblock Generative artificial intelligence in robotic manipulation: A survey.
\newblock {\em arXiv preprint arXiv:2503.03464}, 2025.

\bibitem{wang2024large}
Jiaqi Wang, Enze Shi, Huawen Hu, Chong Ma, Yiheng Liu, Xuhui Wang, Yincheng Yao, Xuan Liu, Bao Ge, and Shu Zhang.
\newblock Large language models for robotics: Opportunities, challenges, and perspectives.
\newblock {\em Journal of Automation and Intelligence}, 2024.

\bibitem{pawar2024survey}
Siddhesh Pawar, Junyeong Park, Jiho Jin, Arnav Arora, Junho Myung, Srishti Yadav, Faiz~Ghifari Haznitrama, Inhwa Song, Alice Oh, and Isabelle Augenstein.
\newblock Survey of cultural awareness in language models: Text and beyond.
\newblock {\em arXiv preprint arXiv:2411.00860}, 2024.

\bibitem{mahadevan2024generative}
Karthik Mahadevan, Jonathan Chien, Noah Brown, Zhuo Xu, Carolina Parada, Fei Xia, Andy Zeng, Leila Takayama, and Dorsa Sadigh.
\newblock Generative expressive robot behaviors using large language models.
\newblock In {\em Proceedings of the 2024 ACM/IEEE International Conference on Human-Robot Interaction}, pages 482--491, 2024.

\bibitem{ren2025touched}
Qiaoqiao Ren and Tony Belpaeme.
\newblock Touched by chatgpt: Using an llm to drive affective tactile interaction.
\newblock {\em arXiv preprint arXiv:2501.07224}, 2025.

\bibitem{cekaite2020towards}
Asta Cekaite and Lorenza Mondada.
\newblock Towards an interactional approach to touch in social encounters.
\newblock In {\em Touch in Social Interaction}, pages 1--26. Routledge, 2020.

\bibitem{willemse2019social}
Christian~JAM Willemse and Jan~BF Van~Erp.
\newblock Social touch in human--robot interaction: Robot-initiated touches can induce positive responses without extensive prior bonding.
\newblock {\em International journal of social robotics}, 11(2):285--304, 2019.

\bibitem{yohanan2012role}
Steve Yohanan and Karon~E MacLean.
\newblock The role of affective touch in human-robot interaction: Human intent and expectations in touching the haptic creature.
\newblock {\em International Journal of Social Robotics}, 4:163--180, 2012.

\bibitem{ren2024tactile}
Qiaoqiao Ren and Tony Belpaeme.
\newblock Tactile interaction with social robots influences attitudes and behaviour.
\newblock {\em International Journal of Social Robotics}, 16(11):2297--2317, 2024.

\bibitem{hertenstein2006touch}
Matthew~J Hertenstein, Dacher Keltner, Betsy App, Brittany~A Bulleit, and Ariane~R Jaskolka.
\newblock Touch communicates distinct emotions.
\newblock {\em Emotion}, 6(3):528, 2006.

\bibitem{janssens2024integrating}
Ruben Janssens, Pieter Wolfert, Thomas Demeester, and Tony Belpaeme.
\newblock Integrating visual context into language models for situated social conversation starters.
\newblock {\em IEEE Transactions on Affective Computing}, 2024.

\end{thebibliography}

\end{document}